\documentclass[a4shrink,portrait,final]{baposter}

\usepackage{times}
\usepackage{calc}
\usepackage{graphicx}
\usepackage{amsmath}
\usepackage{amssymb}
\usepackage{relsize}
\usepackage{multirow}
\usepackage{bm}

\usepackage{graphicx}
\usepackage{multicol}

\usepackage{pgfbaselayers}
\pgfdeclarelayer{background}
\pgfdeclarelayer{foreground}
\pgfsetlayers{background,main,foreground}

\usepackage{helvet}
\usepackage{bookman}
\usepackage{palatino}

\newtheorem{theorem}{Theorem}
\newtheorem{corollary}{Corollary}

\makeatletter
\def\thebibliography#1{\section*{\@mkboth
  {}{}}\list
  {[\arabic{enumi}]}{\settowidth\labelwidth{[#1]}\leftmargin\labelwidth   
\advance\leftmargin\labelsep
\usecounter{enumi}}
\def\newblock{\hskip .11em plus .33em minus .07em}
\sloppy\clubpenalty4000\widowpenalty4000
\sfcode`\.=1000\relax}
\makeatother

\selectcolormodel{cmyk}

\graphicspath{{images/}}

\begin{document}

\title{Language Recognition by Generalized Quantum Finite Automata \\ with Unbounded Error\thanks{This work was
	partially supported by the Scientific and Technological Research Council of Turkey 
	(T\"{U}B\.ITAK) with grant 108142 and the Bo\u{g}azi\c{c}i University Research Fund with grant 08A102.}}
\author{Abuzer Yakary{\i}lmaz\ ~~~~~~ A. C. Cem Say 
	\\
	Bo\u{g}azi\c{c}i University, Department of Computer Engineering,\\ Bebek 34342 \.{I}stanbul, Turkey \\
	\\
	\texttt{ \{say,abuzer\}@boun.edu.tr }
}
\date{March 29, 2009}
\maketitle

\begin{center}

\begin{minipage}{0.70\textwidth}
	\large
	Research on theoretical models of practically implementable quantum
computers has focused on variants of the
quantum finite automaton (QFA). The bounded-error language recognition
capabilities of various alternative
QFA types have been analyzed and compared with their classical
counterparts. In this paper, we examine the
computational power of QFA's in the unbounded error setting. Extending
previous work \cite{YS09D},
we show that all one-way QFA models that are at least as general as
Kondacs-Watrous QFA's (KWQFA's) \cite{KW97}
are equivalent in power to classical probabilistic finite automata in
this setting.
Unlike their probabilistic counterparts, allowing the tape head to stay put for
some steps during its traversal of the input does enlarge the class of
languages recognized by such QFA's
with unbounded error.

A Nayak QFA (NQFA) \cite{Na99} undergoes three
operations in each step of the traversal of its input tape: First, its
state vector evolves
according to the unitary transformation dictated by the scanned
symbol. Then, it undergoes the projective measurement
associated with the same symbol. Finally, another measurement is
performed to see whether the machine has accepted, rejected, or not
halted yet. The computation continues with the next tape symbol
only if this measurement yields the result that the machine has not
halted yet. A KWQFA is a NQFA where the first measurement described
above is just the identity operator.

\begin{theorem}
       \label{theorem:GQFAToGPFA}
       Let $ \mathcal{G}_{1} $ be a NQFA with $ n $ states and $ f_{\mathcal{G}_{1}}:\Sigma \rightarrow [0,1] $ be
       its acceptance probability function.
       Then, there exists a generalized probabilistic finite automaton 
       (GPFA) \cite{Tu69} $ \mathcal{G}_{2} $ with $ O(n^{2}) $ states such that
       $ f_{\mathcal{G}_{1}}(w)=f_{\mathcal{G}_{2}}(w) $ for all $ w \in \Sigma^{*} $.
\end{theorem}
Theorem \ref{theorem:GQFAToGPFA} establishes that every language
recognized with unbounded error
by an NQFA is stochastic.  Combining this with the fact \cite{YS09D}
that every stochastic language can be recognized by a KWQFA, we obtain

\begin{corollary}
       \label{corollary:GQFAStochaticLanguage}
       The class of languages recognized with unbounded error by NQFA's
equals the class of stochastic languages.
\end{corollary}

Several other one-way QFA models (like \cite{Pa00,BMP03}, and the one-way
version of the machines of \cite{AW02},) that generalize the KWQFA have
appeared in the literature. In the bounded-error case, some of these
generalized machines recognize more languages than the KWQFA. We claim
that the classes of languages recognized with unbounded error by all
these automata are identical to each other.

We demonstrate this fact for one of the most general models, namely,
the quantum finite automaton with control language (QFC) \cite{BMP03}, the
proofs for the other variants are similar. For any QFC $ \mathcal{M}
$, there exists a GPFA that computes exactly the same
acceptance probability function as $ \mathcal{M} $, and for any KWQFA
$ \mathcal{M}_{1} $, there exists a QFC that computes
the same acceptance probability function as $ \mathcal{M}_{1} $
\cite{LQ08}. Therefore, QFC's recognize all and only the stochastic
languages with unbounded error.

Regardless of the specifics of their definitions, the two--way
versions of any of these models will have to contain two--way KWQFA's
as
specimens, and the fact \cite{YS09D} that even ``1.5--way" KWQFA's can
recognize nonstochastic languages, combined with
our Theorem \ref{theorem:GQFAToGPFA}, show that such an additional
capability would enlarge
the class of languages recognized by all these QFA's with unbounded error.
\newline
\newline
\newline
\end{minipage}

\end{center}

\definecolor{silver}{cmyk}{0,0,0,0.3}
\definecolor{yellow}{cmyk}{0,0,0.9,0.0}
\definecolor{reddishyellow}{cmyk}{0,0.22,1.0,0.0}
\definecolor{black}{cmyk}{0,0,0.0,1.0}
\definecolor{darkYellow}{cmyk}{0,0,1.0,0.5}
\definecolor{darkSilver}{cmyk}{0,0,0,0.1}

\definecolor{lightyellow}{cmyk}{0,0,0.3,0.0}
\definecolor{lighteryellow}{cmyk}{0,0,0.1,0.0}
\definecolor{lighteryellow}{cmyk}{0,0,0.1,0.0}
\definecolor{lightestyellow}{cmyk}{0,0,0.05,0.0}

\typeout{Poster Starts}
\background{
  \begin{tikzpicture}[remember picture,overlay]%
    \draw (current page.north west)+(-2em,2em) node[anchor=north west] 
    {\includegraphics[height=1.1\textheight]{silhouettes_background}};
  \end{tikzpicture}%
}

\newlength{\leftimgwidth}
\begin{poster}%
  {
  grid=no,
  colspacing=1em,
  bgColorOne=lighteryellow,
  bgColorTwo=lightestyellow,
  borderColor=reddishyellow,
  headerColorOne=yellow,
  headerColorTwo=reddishyellow,
  headerFontColor=black,
  boxColorOne=lightyellow,
  boxColorTwo=lighteryellow,
  textborder=roundedleft,
  eyecatcher=no,
  headerborder=open,
  headerheight=0.10\textheight,
  headershape=roundedright,
  headershade=plain,
  headerfont=\Large\textsf, 
  boxshade=plain,
  background=shade-tb,
  background=plain,
  linewidth=2pt
  }
  {\includegraphics[width=10em]{D1077}} 
	{\sf 
		Language Recognition by Generalized Quantum Finite \\ Automata with Unbounded Error
	}
  {\sf \\ Abuzer Yakary{\i}lmaz (abuzer@boun.edu.tr) and A. C. Cem Say (say@boun.edu.tr) \\
  Bo\u{g}azi\c{c}i University, Department of Computer Engineering, Bebek 34342 \.{I}stanbul, Turkey
  }

 \tikzstyle{light shaded}=[top color=baposterBGtwo!30!white,bottom color=baposterBGone!30!white,shading=axis,shading angle=30]


    %
    \newcommand{\colouredcircle}[1]{%
      \tikz{\useasboundingbox (-0.2em,-0.32em) rectangle(0.2em,0.32em); \draw[draw=black,fill=baposterBGone!80!black!#1!white,line width=0.03em] (0,0) circle(0.18em);}}
\large{
  \headerbox{Main Result}{name=Introduction,column=0,row=0,span=3}{
\LARGE{
We show that all one-way QFA models that are at least as general as Kondacs-Watrous QFA's (KWQFA's) \cite{KW97}
are equivalent in power to classical probabilistic finite automata in the unbounded error setting. 
Unlike their probabilistic counterparts, allowing the tape head to stay put for some steps during its 
traversal of the input does enlarge 
the class of languages recognized by such QFA's with unbounded error.
}
}
  \headerbox{Nayak QFA}{name=NayakQFA,column=0,below=Introduction,span=3}{

A Nayak QFA (NQFA) \cite{Na99} undergoes three operations in each step of the traversal of its input tape: First, its
state vector evolves according to the unitary transformation dictated by the scanned
symbol. Then, it undergoes the projective measurement associated with the same symbol. Finally, another measurement is
performed to see whether the machine has accepted, rejected, or not halted yet. 
The computation continues with the next tape symbol only if this measurement yields the result that the machine has not
halted yet. A KWQFA is a NQFA where the first measurement described above is just the identity operator.

\begin{theorem}
       \label{theorem:GQFAToGPFA-poster}
       Let $ \mathcal{G}_{1} $ be a NQFA with $ n $ states and $ f_{\mathcal{G}_{1}}:\Sigma \rightarrow [0,1] $ be
       its acceptance probability function.
       Then, there exists a generalized probabilistic finite automaton 
       (GPFA) \cite{Tu69} $ \mathcal{G}_{2} $ with $ O(n^{2}) $ states such that
       $ f_{\mathcal{G}_{1}}(w)=f_{\mathcal{G}_{2}}(w) $ for all $ w \in \Sigma^{*} $.
\end{theorem}
Theorem \ref{theorem:GQFAToGPFA-poster} establishes that every language recognized with unbounded error
by an NQFA is stochastic.  Combining this with the fact \cite{YS09D}
that every stochastic language can be recognized by a KWQFA, we obtain 

\begin{corollary}
       \label{corollary:GQFAStochaticLanguage}
       The class of languages recognized with unbounded error by NQFA's
equals the class of stochastic languages.
\end{corollary}
}

  \headerbox{Other QFA's}{name=OtherQFA,column=0,below=NayakQFA,span=3}{
Several other one-way QFA models (like \cite{Pa00,BMP03}, and the one-way
version of the machines of \cite{AW02},) that generalize the KWQFA have
appeared in the literature. In the bounded-error case, some of these
generalized machines recognize more languages than the KWQFA. We claim
that the classes of languages recognized with unbounded error by all
these automata are identical to each other.

We demonstrate this fact for one of the most general models, namely,
the quantum finite automaton with control language (QFC) \cite{BMP03}, the
proofs for the other variants are similar. For any QFC $ \mathcal{M}
$, there exists a GPFA that computes exactly the same
acceptance probability function as $ \mathcal{M} $, and for any KWQFA
$ \mathcal{M}_{1} $, there exists a QFC that computes
the same acceptance probability function as $ \mathcal{M}_{1} $
\cite{LQ08}. Therefore, QFC's recognize all and only the stochastic
languages with unbounded error.
}

  \headerbox{two--way QFA}{name=TwoWayQFA,column=0,below=OtherQFA,span=3}{
Regardless of the specifics of their definitions, the two--way
versions of any of these models will have to contain two--way KWQFA's as
specimens, and the fact \cite{YS09D} that even ``1.5--way" KWQFA's can
recognize nonstochastic languages, combined with
our Theorem \ref{theorem:GQFAToGPFA-poster}, show that such an additional
capability would enlarge
the class of languages recognized by all these QFA's with unbounded error.
}

  \headerbox{References}{name=References,column=0,span=3,below=TwoWayQFA}{
    \smaller
    \vspace{-14pt}
    \bibliographystyle{plain}
	\bibliography{YakaryilmazSay}
}
} 
\end{poster}%
\end{document}